# Dynamic Tuning of Single-Photon Emission in Monolayer $WSe_2$ via Localized Strain Engineering

Yi Yu,[∇] Junyu Ge,[∇] Manlin Luo, In Cheol Seo, Youngmin Kim, John J. H. Eng, Kunze Lu, Tian-Ran Wei, Seok Woo Lee, Weibo Gao, Hong Li,* and Donguk Nam*

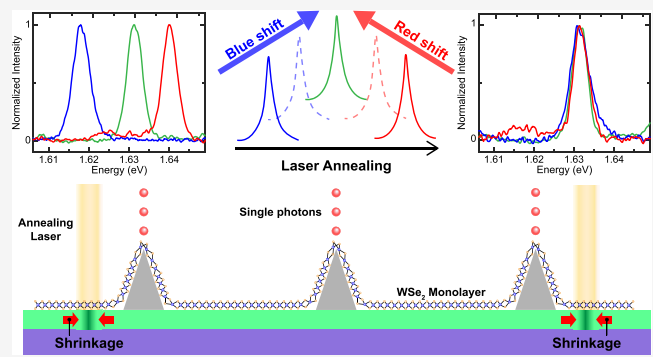

**ABSTRACT:** Two-dimensional (2D) materials have emerged as promising candidates for next-generation integrated single-photon emitters (SPEs). However, significant variability in the emission energies presents a major challenge in producing identical single photons from different 2D SPEs, which may become crucial for practical quantum applications. Although various approaches to dynamically tuning the emission energies of 2D SPEs have been developed to address the issue, the practical solution to matching multiple individual 2D SPEs is still scarce. In this work, we demonstrate precise emission energy tuning of individual SPEs in a $WSe_2$ monolayer. Our approach utilizes localized strain fields near individual SPEs, which we control by adjusting the volume of a stressor layer through laser annealing. This technique allows continuous emission energy tuning of up to 15 meV while maintaining the qualities of SPEs. Additionally, we showcase the precise spectral alignment of three distinct SPEs in a single $WSe_2$ monolayer to the same wavelength.

**KEYWORDS:** single-photon emitters, $WSe_2$ monolayer, dynamic wavelength tunability, strain engineering, laser annealing

Integrating single photon emitters (SPEs) with linear optics to develop quantum photonic integrated circuits (QPICs) has attracted attention as a promising avenue for various quantum applications, including quantum communication[1−3] and quantum information processing.[3−7] Toward this aim, the deterministic and high-efficiency coupling of SPEs with on-chip waveguides is crucial, while conventional SPEs in three-dimensional (3D) materials such as diamond, SiC, or III−V semiconductors often suffer from low photon extraction rates and random emitter positions.[8−11] Recently, transition-metal dichalcogenide (TMD) two-dimensional (2D) materials, especially $WSe_2$ monolayer, have emerged as promising hosts for SPEs that can potentially address the limitations of SPEs in 3D materials.[8−11] For example, the reduced dielectric screening in 2D materials allows a superior photon extraction rate compared to 3D materials. Also, the deterministic creation of SPEs is possible in $WSe_2$ monolayer by harnessing various techniques such as patterned nanopillars,[12,13] nanoindentation,[14,15] and electron beam irradiation,[16,17] enabling high-efficiency coupling of 2D SPEs into on-chip waveguides.[18] Besides, spin angular momentum encoded SPEs and electrically driven SPEs have been demonstrated in $WSe_2$ monolayer, providing a range of diverse functionalities.[19,20]

To achieve QPICs based on 2D SPEs with scalable computational capacity, it is crucial to develop multiple 2D SPEs generating identical single photons. Unfortunately, SPEs in $WSe_2$ monolayers typically exhibit emission energies varying across a large range of 150 meV, making it challenging to produce identical single photons from multiple SPEs. This intrinsic variability in the emission energy is typically attributed to the extreme sensitivity of $WSe_2$ monolayers to inhomogeneous strain in the surrounding environments, which varies the confinement potential of the SPE defect states.[11]

Recently, researchers have demonstrated various postfabrication methods to adjust the emission energies of SPEs in $WSe_2$ monolayers through strain[21,22] and electric field[23,24] engineering, aiming at aligning the emission energies of multiple SPEs. However, existing dynamic tuning methods rely heavily on electric fields and lack the ability to tune each SPE in the 2D flake individually.[25] Consequently, the production of single photons at the same wavelength from multiple SPEs in $WSe_2$ monolayers remains challenging.

In this work, we demonstrate on-demand emission energy tuning of individual SPEs in $WSe_2$ monolayers through dynamic strain engineering. Our method harnesses localized laser annealing to induce controllable shrinkage of the physical volume of an SU-8-based stressor layer underneath WSe$_2$ monolayers. The amount of volume shrinkage in the stressor layer is directly correlated to the amount of temperature rise due to laser annealing, enabling the precise control of local strain fields near individual SPEs upon laser annealing. Our experiments show that the SPE purity and stability are preserved during continuous tuning of SPE emission energy. We also showcase the high-precision spectral alignment of multiple distinct SPEs in the same WSe$_2$ monolayer with considerably different initial emission energies.

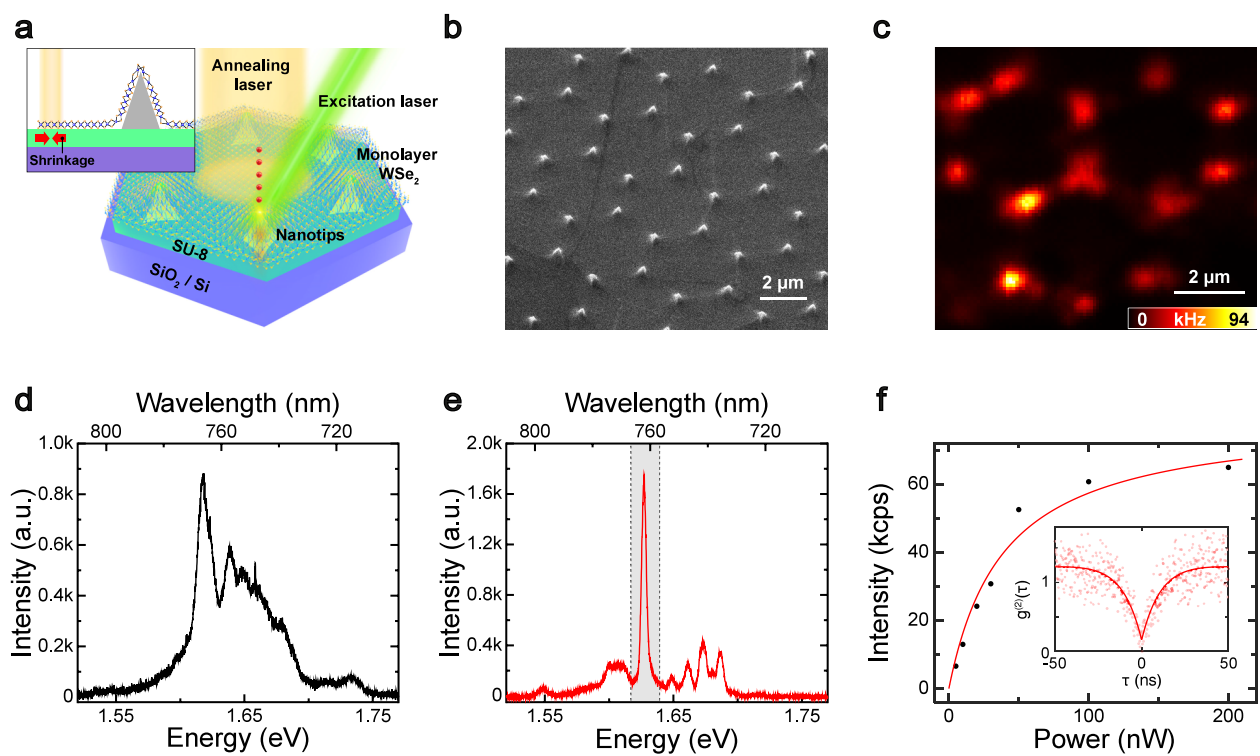

**Figure 1.** (a) Schematic illustration showing how dynamic emission energy tuning of individual SPEs is achieved in a WSe$_2$ monolayer. The WSe$_2$ monolayer is integrated on nanotip arrays with the SU-8 layer. SPEs are generated at the tip positions emitting single photons upon laser excitation (green). The annealing laser (yellow) is focused on the SU-8 layer near the nanotips, controlling the emission energy of single photons. The inset is a cross-section illustration of the device showing the nanotip peaks are not covered by the SU-8 layer. The 10 nm Al$_2$O$_3$ space layer is not shown. (b) Scanning electron microscopy image of the fabricated device, showing the WSe$_2$ monolayer conformally covering the nanotips. (c) Photoluminescence mapping of the WSe$_2$ monolayer emission integrated from 700 to 850 nm, under 532 nm continuous-wave laser excitation. (d and e) WSe$_2$ monolayer emission spectrum from a flat region and from a nanotip, respectively. a.u., arbitrary unit. (f) Excitation power dependence of the integrated intensity of the filtered sharp emission line in the spectral window marked with two dashed lines in panel e, revealing saturation behavior at high excitation powers. The red line is a fitted curve using an empirical saturation model. The inset shows the second-order correlation for the same sharp emission line. The red line indicates a fitted curve using a double exponential model. The filtering was done using a tunable bandpass filter with a 10 nm bandwidth to collect only the sharp emission line centered at 762 nm.

Figure 1a shows the schematic illustration of our proposed device for continuous tuning of emission energy from strain-induced SPEs in WSe$_2$. Our device consists of nanotip arrays to facilitate the deterministic creation of SPEs in WSe$_2$ as well as an SU-8 layer that is firmly adhered to the flat region of the WSe$_2$ monolayer. The SU-8 layer works as a stressor layer owing to its ability to change its physical volume upon localized laser annealing. The local shrinkage of the SU-8 layer adjusts the strain applied to the SPEs in the WSe$_2$ monolayer at the nanotips, leading to effective SPE emission energy tuning.

For the device fabrication, we first defined hexagonal Cr nanotip arrays on a SiO$_2$ substrate using nanosphere lithography (NSL) with polystyrene (PS) spheres (see Supporting Information S1 for details on the NSL process). Next, a 10 nm Al$_2$O$_3$ spacer layer was deposited via atomic layer deposition (ALD). This space layer serves a dual purpose: preventing SPE quenching by the Cr nanotips and mitigating spectral diffusion caused by surface charge noise on the substrate.[26] Subsequently, an SU-8 layer was spin-coated onto the spacer layer, followed by a soft bake for solidification. As shown in the inset to Figure 1a, the nanotip peaks are exposed without being covered by the SU-8 layer. This is important for preventing the charge environment in the SU-8 layer from disturbing the properties of SPEs at the nanotips. Following this, we subjected the SU-8 layer to ultraviolet (UV) exposure and a postexposure bake. The acid catalyst generated under the UV exposure renders the SU-8 layer the ability to shrink upon laser annealing.[27] Atomic force microscopy (AFM) mapping of the device reveals nanotip heights and diameters of 230 ± 10 nm and 590 ± 10 nm, respectively. A thin WSe$_2$ film was mechanically exfoliated onto a commercial PDMS stamp and characterized by optical contrast and Raman spectroscopy to confirm its monolayer nature (see Supporting Information S2). The WSe$_2$ monolayer was then dry transferred onto the prepared substrate with nanotip arrays. Figure 1b presents a scanning electron microscope (SEM) image of the fabricated device, showing that the WSe$_2$ monolayer conformally covers the nanotips and firmly adheres to the SU-8 layer. Finally, the sample was mounted on a low-vibration closed-cycle cryostat for optical characterization.

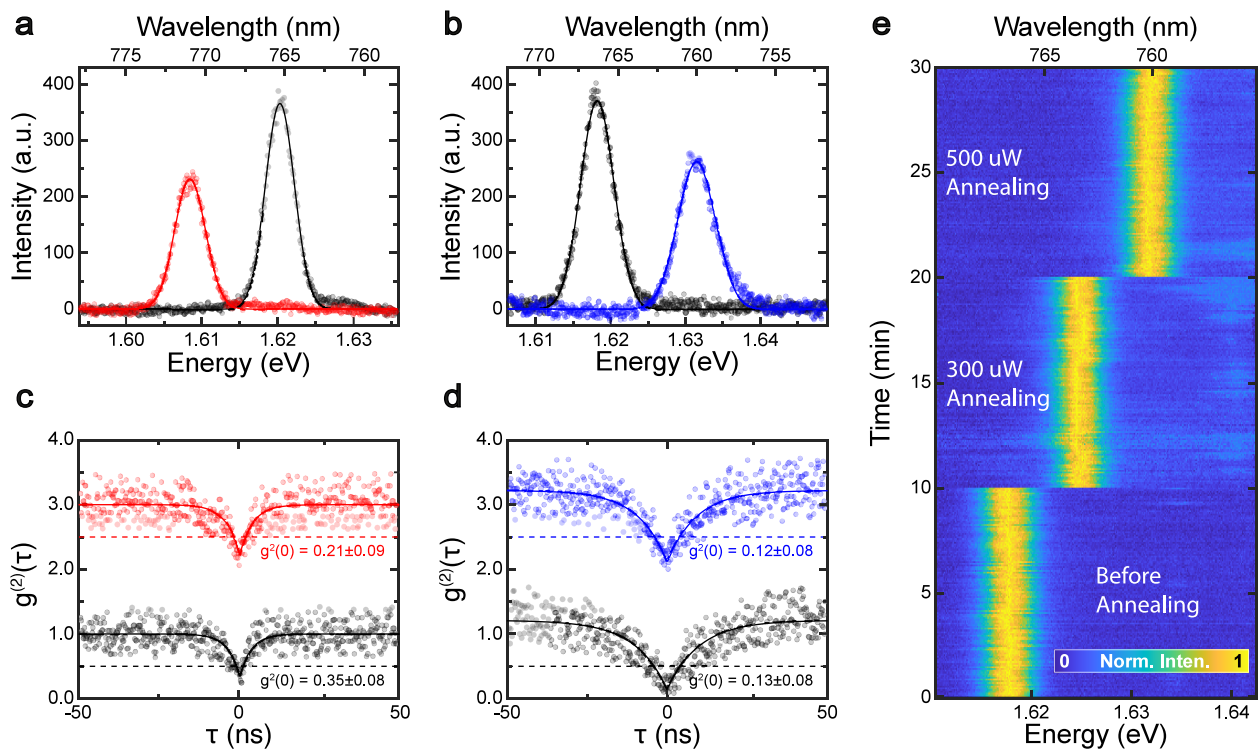

**Figure 2.** (a and b) SPE emission spectra before (black) and after (red or blue) 500 μW laser annealing for S1 and S2, respectively. All spectra were obtained under 30 nW laser excitation. (c and d) Second-order correlation before (black) and after (red or blue) 500 μW laser annealing for S1 and S2, respectively. The uncertainty represents the standard deviation obtained from the fitting. (e) Time-resolved spectral acquisition for S2 before (bottom), after 300 μW (middle), and after 500 μW (top) laser annealing. Each panel consists of 200 consecutive spectral acquisitions of 3 s each. Each panel is normalized individually.

We conducted photoluminescence (PL) measurements on the fabricated device at a low temperature of 4 K (see Methods for details on the experimental setup). We first performed PL mapping of the device under a 532 nm continuous-wave (CW) laser excitation. Figure 1c shows the 2D mapping of PL intensity integrated between 700 and 850 nm. Notably, strong PL emission from WSe$_2$ is observed at the nanotips, whereas significantly weaker emission is observed from the flat region, indicating efficient funneling of the excitons in the WSe$_2$ monolayer toward the nanotips due to the large strain gradient near the nanotips.[12]

Figure 1d and 1e show PL spectra at the flat region and the nanotip, respectively. Due to the relatively weaker emission from the flat region, we used different excitation powers of 10 μW and 100 nW in the flat region and on the nanotip, respectively. The flat region produces a broadband emission spanning from 1.58 to 1.69 eV (Figure 1d), consistent with the defect broadband emission reported in previous works.[12,17,21,22,28] Additionally, two relatively weak peaks near 1.74 and 1.71 eV are observed, which can be attributed to bright exciton and trion emissions.[21,28] Conversely, the spectrum measured at the nanotip exhibits sharp and isolated emission lines within the defect energy range (Figure 1e). Figure 1f presents the excitation-power-dependent intensity of the sharp emission line filtered in the spectral window marked with two vertical dashed lines in Figure 1e, revealing a saturation behavior at high excitation powers. The measured data points are fitted to the following empirical saturation model:

$$I = \frac{I_{sat}P}{P + P_{sat}}$$

where $I$ and $P$ are the emission intensity and excitation power, respectively, while $I_{sat}$ and $P_{sat}$ are the saturation intensity and saturation power, respectively. The saturation intensity $I_{sat}$ of 40 kcps at $P_{sat}$ of 39.3 nW is extracted. The strong saturation indicates the defect-related character of these sharp emission lines, which is widely attributed to the hybridization between strain-confined intervalley dark exciton states and localized defect states.[29] Further investigation of the anticipated single-photon nature is obtained through the photon statistics study. The inset to Figure 1f depicts the second-order correlation, $g^2(\tau)$, obtained from the filtered emission line in Figure 1e without background subtraction. The measured data can be fitted to a double exponential model function:

$$g^2(\tau) = 1 - b[(1+a)e^{-|\tau|/T_1} - ae^{-|\tau|/T_2}]$$

where $T_1$ and $T_2$ are decay lifetime and bunching time, respectively (see Supporting Information S3). The $g^2(\tau)$ data dips well below 0.5 at $\tau = 0$ reveals the SPE nature of the isolated emission line. The extracted $T_1$ of 10.95 ± 1.1 ns is close to the previously reported values.[13]

Upon focusing a high-power laser onto the SU-8 layer within a small area, as illustrated in Figure 1a, the local temperature in the SU-8 layer increases. The temperature at the center of the annealed area can reach around 480 K by focusing a 500 μW 532 nm CW laser within a laser spot of 1 μm diameter (see Supporting Information S4). Temperatures below 500 K can induce volume shrinkage of around 8% in the SU-8 layer due to cross-links,[27] while higher temperatures can cause shrinkage above 10% due to oxygen and hydrogen dissociation.[30]

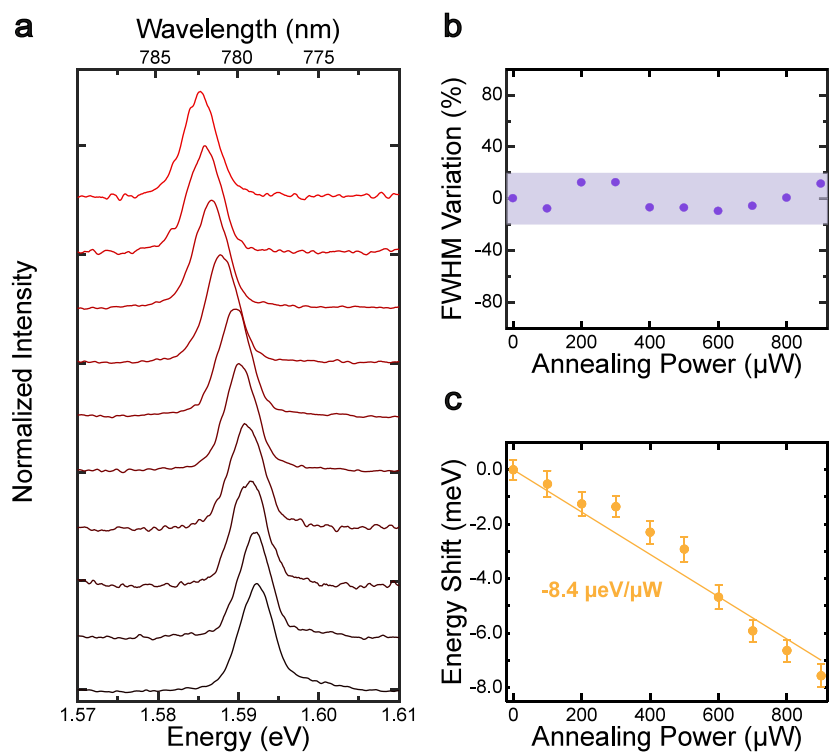

**Figure 3.** (a) SPE emission spectra measured prior to laser annealing (bottom) and after laser annealing at varying powers from 100 μW (second from the bottom) to 900 μW (top) in 100 μW increments. Annealing is conducted for 30 s for all annealing powers. (b) Variation of the emission line full width at half-maximum (fwhm) with respect to the initial emission line as a function of annealing power. The shaded area represents the line width range within 20% of the initial line width. (c) Emission energy shift as a function of annealing power. A tuning rate of −8.4 μeV/μW can be extracted from the fitting line. The error bar represents the standard deviation of the peak energy obtained from the spectrum fitting.

To experimentally demonstrate the SPE emission energy tuning via laser annealing, we measured SPE emission spectra before and after localized laser annealing at different powers in steps of 100 μW. The SPE spectrum before annealing was first measured under the 532 nm CW laser excitation. The same laser was then positioned at the annealing spot located 2 μm away from the nanotip by utilizing a fine-resolution dual-axis Galvo scanner. The 2 μm distance was selected to avoid damaging the SPEs from the high-power laser with a laser spot of around 1 μm diameter. Laser annealing was then performed with a 30-s annealing duration to ensure a stable SU-8 shrinkage at the corresponding powers.[31] After the annealing, the laser was returned to the nanotip region for the SPE spectrum acquisition. To ensure that the SPE emission is not influenced by potential temperature rise caused by laser annealing, a minimum time interval of 1 min is maintained between the laser annealing and the SPE spectrum measurement.

Figure 2a and 2b show the spectra before (black) and after 500 μW (red and blue) laser annealing for two different SPEs, S1 and S2, respectively. The results clearly demonstrate the effective tuning of the SPE emission energy through laser annealing. Interestingly, the magnitude and direction of emission energy shifts vary among SPEs. For example, S1 exhibits a redshift of 12.6 meV, suggesting increased tensile strain experienced at the SPE site after laser annealing, while S2 displays a blueshift of 15 meV, indicative of decreased tensile strain. This disparity likely arises from the different strain-tuning responses of SPEs at various positions of the nanotips, induced by the stressor-layer shrinkage during laser annealing.[21] The angle between the strain-tuning direction and the SPE dipole orientation could contribute to different shift directions,[32] while the magnitude of the strain variation could result in different shift magnitudes. Nevertheless, once the shift direction for a given SPE is identified (e.g., through a brief preannealing step), it remains consistent across all annealing powers tested (100 μW−1000 μW). This reliable behavior enables us to dynamically tune SPE emission energies with confidence.

To investigate whether the SPE qualities are preserved, we evaluate the SPE purity and the spectral stability before and after the annealing. The red and blue data points and fitted curves in Figure 2c and 2d are the $g^2(\tau)$ function obtained after the 500-μW laser annealing for S1 and S2, respectively. The measured data before the annealing is also shown in black for comparison. The majority of measured SPEs, represented by S2 in Figure 2D, showed the preserved purity after the annealing, which is evidenced by comparable $g^2(0)$ values before and after the annealing. In contrast, some SPEs exhibited an enhanced purity upon laser annealing. For instance, S1 shows a 40% reduction in the $g^2(0)$ value. However, there is no clear correlation between the spectrum shift direction and the SPE purity enhancement. The exact origin of the enhanced SPE purity requires further theoretical and experimental studies. Figure 2e displays the spectral jitter recorded for S2 before and after the laser annealing at two annealing powers of 300 μW and 500 μW. While the consistent spectrum shift caused by laser annealing at different powers can be clearly seen, it is also noteworthy that the jittering remains comparable before and after the laser annealing.

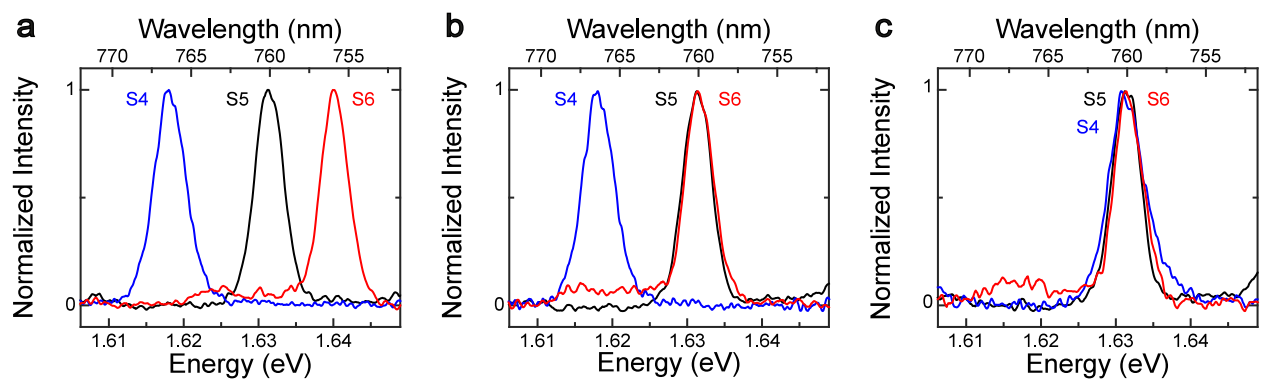

**Figure 4.** (a) Normalized SPE emission spectra of three SPEs before the spectral matching. (b) Spectra after aligning S6 to S5. (c) Spectra after aligning S4 to S5 and S6.

To provide a full picture of SPE emission energy tuning via laser annealing, we recorded the emission spectra of another SPE S3 after conducting the laser annealing at increasing annealing powers ranging from 100 μW to 900 μW in 100 μW increments (Figure 3a). The spectra before and after laser annealing with increasing power are presented from bottom to top. The extracted emission full width at half-maximum (fwhm) variation falls within 20% of the initial line width throughout the laser annealing process, showing good spectral stability (Figure 3b). The consistent line width throughout the laser annealing process further confirms that the SPE emission energy tuning is not due to the temperature increase induced by the laser annealing, which will otherwise broaden the emission line width.[16,33] The emission energy shift of S3 as a function of laser annealing powers is shown in Figure 3c. A continuous redshift with a tuning rate of −8.4 μeV/μW was observed. The energy shift exhibits a linear relationship with the laser annealing power with a sub-meV tuning precision for the 100 μW increment step. The tuning precision is expected to be further enhanced by employing a smaller laser annealing power increment step.

Finally, we present the spectral matching of three SPEs from the same WSe$_2$ monolayer with initially different emission energies. Figure 4a presents the initial spectra of three SPEs, S4 (blue), S5 (black), and S6 (red), exhibiting emission lines centered at 1.618, 1.631, and 1.640 eV, respectively. The discrepancy in emission energies likely stems from the nonuniform strain distribution induced during the dry-transfer process.[9] To match the emission energies of S5 and S6, the laser annealing was applied to S6 with annealing powers starting from 100 μW in 100 μW increments, while the spectrum of S6 was monitored after each annealing step. Figure 4b presents the emission spectra of three SPEs obtained after applying 500 μW laser annealing to S6. The emission line of S6 is red-shifted to align with S5 with sub-meV precision, while the retaken emission spectra of S4 and S5 remain the same. Subsequently, the laser annealing was also performed on S4 to align its emission energy with S5 and S6. After 500 μW laser annealing was applied to S4, the S4 emission line is blueshifted to align with S5 and S6 emission lines (Figure 4c). The initial discrepancy in emission energies of up to 22 meV was largely eliminated, leading to an energy difference of less than 0.5 meV among the three distinct SPEs. This energy difference can be further reduced by increasing the annealing laser power at a finer step to induce a smaller spectral shift of SPEs.

The capability to align different SPE emission lines demonstrates the precise energy tuning of individual SPE, showing the potential of our laser annealing technique for obtaining identical single photons from distinct SPEs. It is noteworthy that utilizing two different lasers for SPE excitation and laser annealing separately enables annealing of the SU-8 stressor layer while monitoring the spectral shift of SPEs in real time.

In conclusion, we have achieved the localized and precise emission energy tuning of individual SPEs from WSe$_2$ monolayers. The SU-8 stressor layer, coupled with patterned nanotip arrays, provides an efficient platform for fine control of both the spatial position and emission energy of individual SPEs. Through laser annealing with varying powers, we realized continuous emission energy tuning of up to 15 meV, which corresponds to approximately 0.3% of strain tuning at the SPE site.[22] Note that the 0.3% strain tuning is similar to other reported strain tuning values realized by utilizing the shrinkage of SU-8.[34] Remarkably, SPE purity and spectral stability are maintained throughout the laser annealing process, confirming that neither the laser annealing nor the strain adjustment degrades the quality of the SPEs at the nanotips. The localized tunability enables the spectral alignment of distinct SPEs from the same WSe$_2$ monolayer, with a significant initial emission energy difference of up to 22 meV. Although the direction of the emission energy shift is predetermined by each SPE's local strain environment, we anticipate that combining laser annealing with engineered structures, such as nanogaps designed for anisotropic strain and dipole orientation control,[35] could enable fully deterministic, directional tuning of WSe$_2$ SPEs. Our localized, permanent, and precise emission energy tuning technique for WSe$_2$ SPEs opens new avenues for realizing indistinguishable SPEs in 2D materials.

## ■ METHODS

**Optical Setup.** We mounted the sample on a 4K low-vibration closed-cycle cryostat. A microscope objective lens (100 ×, NA = 0.9) focused the continuous-wave (cw) laser (532 nm) within a laser spot of around 1 μm$^2$ and collected the photoluminescence (PL) signal from WSe$_2$ monolayers. For large-area PL mapping, the focused laser was scanned through the area on the sample, with the help of a fine-resolution dual-axis Galvo scanner (Thorlabs GVS002), for both excitation and emission collection. For spectra acquisition, we sent the emission to a high-resolution (0.05 nm) spectrometer with an air-cooled Si array charge-coupled device (Andor). For second-order photon correlation measurement, we sent the spectral-filtered signals to a Hanbury-Brown and Twiss setup consisting of a 50:50 beamsplitter followed by two Si single photon avalanche photodetectors connected to a time-correlated single photon counter module (Picoharp 300). The timing resolution of the system is around 40 ps, mostly limited by the detectors.

## ■ ASSOCIATED CONTENT

### *ⓢ* Supporting Information

The Supporting Information is available free of charge at https://pubs.acs.org/doi/10.1021/acs.nanolett.4c05450.

> Additional device fabrication on Cr nanotip arrays, WSe$_2$ monolayer characterization, SPE $g^2(\tau)$ fitting, thermal distribution simulation results for the SU-8 stressor layer, optical microscopy of the device, and polarization of a characteristic SPE (PDF)

## ■ AUTHOR INFORMATION

### Corresponding Authors

**Donguk Nam** − *Department of Mechanical Engineering, Korea Advanced Institute of Science and Technology (KAIST), Daejeon 34141, Republic of Korea;* orcid.org/0000-0003-4378-205X; Email: dwnam@kaist.ac.kr

**Hong Li** − *School of Electrical and Electronic Engineering and School of Mechanical and Aerospace Engineering, Nanyang Technological University (NTU), Singapore 639798; CINTRA CNRS/NTU/THALES, IRL 3288, Research Techno Plaza, Nanyang Technological University, Singapore 637553;* orcid.org/0000-0002-6975-7787; Email: ehongli@ntu.edu.sg

### Authors

**Yi Yu** − *School of Electrical and Electronic Engineering, Nanyang Technological University (NTU), Singapore 639798*

**Junyu Ge** − *School of Mechanical and Aerospace Engineering, Nanyang Technological University (NTU), Singapore 639798*

**Manlin Luo** − *School of Electrical and Electronic Engineering, Nanyang Technological University (NTU), Singapore 639798*

**In Cheol Seo** − *School of Electrical and Electronic Engineering, Nanyang Technological University (NTU), Singapore 639798; A*STAR Quantum Innovation Centre (Q.InC), National Metrology Centre (NMC), Agency for Science, Technology and Research (A*STAR), Singapore 637145*

**Youngmin Kim** − *School of Electrical and Electronic Engineering, Nanyang Technological University (NTU), Singapore 639798; School of Materials Science and Engineering, Kookmin University, Seoul 02707, Republic of Korea;* orcid.org/0000-0001-7248-2078

**John J. H. Eng** − *School of Physical and Mathematical Sciences, Nanyang Technological University (NTU), Singapore 639798; Institute of Materials Research and Engineering (IMRE), Agency of Science, Technology and Research (A*STAR), Singapore 138634*

**Kunze Lu** − *School of Electrical and Electronic Engineering and School of Physical and Mathematical Sciences, Nanyang Technological University (NTU), Singapore 639798;* orcid.org/0000-0003-1132-1822

**Tian-Ran Wei** − *State Key Laboratory of Metal Matrix Composites, School of Materials Science and Engineering, Shanghai Jiao Tong University, Shanghai 200240, China;* orcid.org/0000-0002-5011-0932

**Seok Woo Lee** − *School of Electrical and Electronic Engineering, Nanyang Technological University (NTU), Singapore 639798;* orcid.org/0000-0003-2459-7174

**Weibo Gao** − *School of Electrical and Electronic Engineering and School of Physical and Mathematical Sciences, Nanyang Technological University (NTU), Singapore 639798;* orcid.org/0000-0003-3971-621X

Complete contact information is available at:
https://pubs.acs.org/10.1021/acs.nanolett.4c05450

### Author Contributions

∇Y.Y. and J.G. contributed equally to this work.

### Notes

The authors declare no competing financial interest.

## ■ ACKNOWLEDGMENTS

The authors acknowledge and thank the Nanyang Nano-Fabrication Centre (N2FC). This research is supported by the National Research Foundation, Singapore, and A*STAR under its Quantum Engineering Programme (NRF2022-QEP2-02-P13) and Ministry of Education Singapore (Grant MOE-T2EP50221-0002). S.W.L. acknowledges the support by the Ministry of Education, Singapore, under AcRF Tier 1 (reference RT8/23).

## ■ REFERENCES

(1) Pelucchi, E.; Fagas, G.; Aharonovich, I.; Englund, D.; Figueroa, E.; Gong, Q.; Hannes, H.; Liu, J.; Lu, C. Y.; Matsuda, N.; Pan, J. W.; Schreck, F.; Sciarrino, F.; Silberhorn, C.; Wang, J.; Jöns, K. D. The Potential and Global Outlook of Integrated Photonics for Quantum Technologies. *Nat. Rev. Phys.* **2022**, *4*, 194−208.

(2) Northup, T. E.; Blatt, R. Quantum Information Transfer Using Photons. *Nat. Photonics* **2014**, *8*, 356−363.

(3) Awschalom, D. D.; Hanson, R.; Wrachtrup, J.; Zhou, B. B. Quantum Technologies with Optically Interfaced Solid-State Spins. *Nat. Photonics* **2018**, *12*, 516−527.

(4) O'Brien, J. L. Optical Quantum Computing. *Science* **2007**, *318*, 1567−1570.

(5) Adcock, J. C.; Vigliar, C.; Santagati, R.; Silverstone, J. W.; Thompson, M. G. Programmable Four-Photon Graph States on a Silicon Chip. *Nat. Commun.* **2019**, *10*, 3528.

(6) Paesani, S.; Borghi, M.; Signorini, S.; Maïnos, A.; Pavesi, L.; Laing, A. Near-Ideal Spontaneous Photon Sources in Silicon Quantum Photonics. *Nat. Commun.* **2020**, *11*, 2505.

(7) Maring, N.; Fyrillas, A.; Pont, M.; Ivanov, E.; Stepanov, P.; Margaria, N.; Hease, W.; Pishchagin, A.; Lemaître, A.; Sagnes, I.; Au, T. H.; Boissier, S.; Bertasi, E.; Baert, A.; Valdivia, M.; Billard, M.; Acar, O.; Brieussel, A.; Mezher, R.; Wein, S. C.; Salavrakos, A.; Sinnott, P.; Fioretto, D. A.; Emeriau, P. E.; Belabas, N.; Mansfield, S.; Senellart, P.; Senellart, J.; Somaschi, N. A Versatile Single-Photon-Based Quantum Computing Platform. *Nat. Photonics* **2024**, *18*, 603−609.

(8) Aharonovich, I.; Englund, D.; Toth, M. Solid-State Single-Photon Emitters. *Nat. Photonics* **2016**, *10*, 631−641.

(9) Azzam, S. I.; Parto, K.; Moody, G. Prospects and Challenges of Quantum Emitters in 2D Materials. *Appl. Phys. Lett.* **2021**, *118*, 240502.

(10) Montblanch, A. R. P.; Barbone, M.; Aharonovich, I.; Atatüre, M.; Ferrari, A. C. Layered Materials as a Platform for Quantum Technologies. *Nat. Nanotechnol.* **2023**, *18*, 555−571.

(11) Atatüre, M.; Englund, D.; Vamivakas, N.; Lee, S. Y.; Wrachtrup, J. Material Platforms for Spin-Based Photonic Quantum Technologies. *Nat. Rev. Mater.* **2018**, *3*, 38−51.

(12) Branny, A.; Kumar, S.; Proux, R.; Gerardot, B. D. Deterministic Strain-Induced Arrays of Quantum Emitters in a Two-Dimensional Semiconductor. *Nat. Commun.* **2017**, *8*, 15053.

(13) Palacios-Berraquero, C.; Kara, D. M.; Montblanch, A. R. P.; Barbone, M.; Latawiec, P.; Yoon, D.; Ott, A. K.; Loncar, M.; Ferrari, A. C.; Atatüre, M. Large-Scale Quantum-Emitter Arrays in Atomically Thin Semiconductors. *Nat. Commun.* **2017**, *8*, 15093.

(14) Rosenberger, M. R.; Dass, C. K.; Chuang, H. J.; Sivaram, S. V.; McCreary, K. M.; Hendrickson, J. R.; Jonker, B. T. Quantum Calligraphy: Writing Single-Photon Emitters in a Two-Dimensional Materials Platform. *ACS Nano* **2019**, *13*, 904−912.

(15) Abramov, A. N.; Chestnov, I. Y.; Alimova, E. S.; Ivanova, T.; Mukhin, I. S.; Krizhanovskii, D. N.; Shelykh, I. A.; Iorsh, I. V.; Kravtsov, V. Photoluminescence Imaging of Single Photon Emitters within Nanoscale Strain Profiles in Monolayer $WSe_2$. *Nat. Commun.* **2023**, *14*, 5737.

(16) Parto, K.; Azzam, S. I.; Banerjee, K.; Moody, G. Defect and Strain Engineering of Monolayer $WSe_2$ Enables Site-Controlled Single-Photon Emission up to 150 K. *Nat. Commun.* **2021**, *12*, 3585.

(17) Xu, D. D.; Vong, A. F.; Lebedev, D.; Ananth, R.; Wong, A. M.; Brown, P. T.; Hersam, M. C.; Mirkin, C. A.; Weiss, E. A. Conversion of Classical Light Emission from a Nanoparticle-Strained $WSe_2$ Monolayer into Quantum Light Emission via Electron Beam Irradiation. *Adv. Mater.* **2023**, *35*, 2208066.

(18) Peyskens, F.; Chakraborty, C.; Muneeb, M.; Van Thourhout, D.; Englund, D. Integration of Single Photon Emitters in 2D Layered Materials with a Silicon Nitride Photonic Chip. *Nat. Commun.* **2019**, *10*, 4435.

(19) Lee, S. J.; So, J. P.; Kim, R. M.; Kim, K. H.; Rha, H. H.; Na, G.; Han, J. H.; Jeong, K. Y.; Nam, K. T.; Park, H. G. Spin Angular Momentum-Encoded Single-Photon Emitters in a Chiral Nanoparticle-Coupled $WSe_2$ Monolayer. *Sci. Adv.* **2024**, *10*, No. eadn7210.

(20) So, J. P.; Kim, H. R.; Baek, H.; Jeong, K. Y.; Lee, H. C.; Huh, W.; Kim, Y. S.; Watanabe, K.; Taniguchi, T.; Kim, J.; Lee, C. H.; Park, H. G. Electrically Driven Strain-Induced Deterministic Single-Photon Emitters in a van Der Waals Heterostructure. *Sci. Adv.* **2021**, *7*, No. eabj3176.

(21) Iff, O.; Tedeschi, D.; Martín-Sánchez, J.; Moczała-Dusanowska, M.; Tongay, S.; Yumigeta, K.; Taboada-Gutiérrez, J.; Savaresi, M.; Rastelli, A.; Alonso-González, P.; Höfling, S.; Trotta, R.; Schneider, C. Strain-Tunable Single Photon Sources in $WSe_2$ Monolayers. *Nano Lett.* **2019**, *19*, 6931−6936.

(22) Kim, H.; Moon, J. S.; Noh, G.; Lee, J.; Kim, J. H. Position and Frequency Control of Strain-Induced Quantum Emitters in $WSe_2$ Monolayers. *Nano Lett.* **2019**, *19*, 7534−7539.

(23) Chakraborty, C.; Goodfellow, K. M.; Dhara, S.; Yoshimura, A.; Meunier, V.; Vamivakas, A. N. Quantum-Confined Stark Effect of Individual Defects in a van Der Waals Heterostructure. *Nano Lett.* **2017**, *17*, 2253−2258.

(24) Mukherjee, A.; Chakraborty, C.; Qiu, L.; Vamivakas, A. N. Electric Field Tuning of Strain-Induced Quantum Emitters in $WSe_2$. *AIP Adv.* **2020**, *10*, No. 075310.

(25) Yu, Y.; Seo, I. C.; Luo, M.; Lu, K.; Son, B.; Tan, J. K.; Nam, D. Tunable Single-Photon Emitters in 2D Materials. *Nanophotonics* **2024**, *13*, 3615−3629.

(26) Li, X.; Shepard, G. D.; Cupo, A.; Camporeale, N.; Shayan, K.; Luo, Y.; Meunier, V.; Strauf, S. Nonmagnetic Quantum Emitters in Boron Nitride with Ultranarrow and Sideband-Free Emission Spectra. *ACS Nano* **2017**, *11*, 6652−6660.

(27) Feng, R.; Farris, R. J. Influence of Processing Conditions on the Thermal and Mechanical Properties of SU8 Negative Photoresist Coatings. *J. Micromech. Microeng.* **2003**, *13*, 80.

(28) Kumar, S.; Kaczmarczyk, A.; Gerardot, B. D. Strain-Induced Spatial and Spectral Isolation of Quantum Emitters in Mono- and Bilayer $WSe_2$. *Nano Lett.* **2015**, *15*, 7567−7573.

(29) Linhart, L.; Paur, M.; Smejkal, V.; Burgdörfer, J.; Mueller, T.; Libisch, F. Localized Intervalley Defect Excitons as Single-Photon Emitters in $WSe_2$. *Phys. Rev. Lett.* **2019**, *123*, 146401.

(30) Park, B. Y.; Taherabadi, L.; Wang, C.; Zoval, J.; Madou, M. J. Electrical Properties and Shrinkage of Carbonized Photoresist Films and the Implications for Carbon Microelectromechanical Systems Devices in Conductive Media. *J. Electrochem. Soc.* **2005**, *152*, J136.

(31) Kim, Y.; Joo, H. J.; Chen, M.; Son, B.; Burt, D.; Shi, X.; Zhang, L.; Ikonic, Z.; Tan, C. S.; Nam, D. High-Precision Wavelength Tuning of GeSn Nanobeam Lasers via Dynamically Controlled Strain Engineering. *Adv. Sci.* **2023**, *10*, 2207611.

(32) Grosso, G.; Moon, H.; Lienhard, B.; Ali, S.; Efetov, D. K.; Furchi, M. M.; Jarillo-Herrero, P.; Ford, M. J.; Aharonovich, I.; Englund, D. Tunable and High-Purity Room Temperature Single-Photon Emission from Atomic Defects in Hexagonal Boron Nitride. *Nat. Commun.* **2017**, *8*, 705.

(33) Luo, Y.; Liu, N.; Li, X.; Hone, J. C.; Strauf, S. Single Photon Emission in $WSe_2$ up 160 K by Quantum Yield Control. *2D Mater.* **2019**, *6*, No. 035017.

(34) Takamura, M.; Hibino, H.; Yamamoto, H. Applying Strain into Graphene by SU-8 Resist Shrinkage. *J. Phys. D: Appl. Phys.* **2016**, *49*, 285303.

(35) So, J. P.; Jeong, K. Y.; Lee, J. M.; Kim, K. H.; Lee, S. J.; Huh, W.; Kim, H. R.; Choi, J. H.; Kim, J. M.; Kim, Y. S.; Lee, C. H.; Nam, S.; Park, H. G. Polarization Control of Deterministic Single-Photon Emitters in Monolayer $WSe_2$. *Nano Lett.* **2021**, *21*, 1546−1554.